\documentclass[%
 reprint,
superscriptaddress,
 amsmath,amssymb,
 aps,
]{revtex4-2}
\usepackage{pdfpages}
\makeatletter
\AtBeginDocument{\let\LS@rot\@undefined}
\makeatother
\usepackage{graphicx}
\usepackage{dcolumn}
\usepackage{comment}
\usepackage{csquotes}
\usepackage{bm}
\usepackage{siunitx}

\usepackage{cleveref}
\crefname{equation}{Eq.}{Eqs.}
\crefname{figure}{Fig.}{Figs.}

\def\Th{^{229}\text{Th}}
\begin{document}
\setlength{\parskip}{0pt}
\setlength{\textfloatsep}{10pt}
\preprint{APS/123-QED}

\title{Collective cavity quantum electrodynamics in solid-state optical clocks}

\author{Karen Mamian}
\affiliation{Institute of Science and Technology Austria, Am Campus 1, 3400 Klosterneuburg, Austria}

\author{Georgy A. Kazakov}
\affiliation{Technische Universität Wien, Atominstitut, Stadionallee 2, 1020 Wien, Austria}
\affiliation{Wolfgang Pauli Institute, Oskar-Morgenstern-Platz 1, A-1090, Vienna, Austria}

\author{Thorsten Schumm}
\affiliation{Technische Universität Wien, Atominstitut, Stadionallee 2, 1020 Wien, Austria}

\author{Charles Roques-Carmes}
\email{crc@ista.ac.at}
\affiliation{Institute of Science and Technology Austria, Am Campus 1, 3400 Klosterneuburg, Austria}

\begin{abstract}
Solid-state frequency standards are generally limited by strong decoherence, rendering conventional interrogation schemes inefficient. The $\Th$ nuclear clock  provides a unique and timely platform for solid-state optical metrology and nuclear cavity quantum electrodynamics (QED), featuring a coherence time many orders of magnitude shorter than the radiative lifetime in current experiments. Here, we propose and analyze three cavity QED-enhanced clock interrogation schemes that turn this timescale mismatch into an advantage, leveraging collective coupling of thorium nuclei to nanophotonic modes to enable fast interrogation and detection despite the long population lifetime. We reveal the central role of collective cooperativity in determining the clock frequency instability, and derive the optimal conditions (power, working-point detuning, thorium density) for clock operation.

\end{abstract}
 
\maketitle

\newpage

Optical atomic clocks realize the most precise measurements of time and frequency, enabling applications ranging from navigation and relativistic geodesy to tests of fundamental physics~\cite{ludlow2015optical,nicholson2015systematic, fortier2026optical}. State-of-the-art optical clocks are based on trapped ions or ensembles of neutral atoms, relying on complex trapping, cooling and laser-stabilization systems~\cite{ludlow2015optical, zheng2024reducing} that limit their scalability. Solid-state platforms, on the other hand, offer a promising route toward robust and transportable optical frequency standards~\cite{hodges2013timekeeping,lourette2026toward}.

Embedding emitters into solid-state hosts exposes them to host-induced frequency shifts, inhomogeneous broadening, and dephasing, inevitably reducing their coherence times far below their radiative lifetimes~\cite{thiel2011rare,ooi2026frequency}.
Collective cavity and waveguide quantum electrodynamics (QED) can nevertheless concentrate the ensemble response into a phase-matched bright mode, enhancing emission into a selected photonic channel and mediating long-range interactions~\cite{plankensteiner2019enhanced,lei2023many,junker2012cooperative,macovei2026phase,kong2016stopping,pak2022long,lukin2025mesoscopic, auffeves2010controlling, zhong2017nanophotonic, nickerson2018collective}. In particular, this physics underlies proposals for superradiant lasers and active optical frequency standards~\cite{meiser2009prospects,Zhang2023,matusko2024superradiant, kazakov2022ultimate, su2016controllable, koppenhofer2023squeezed, koppenhofer2022dissipative}.
 
Recently, laser excitation of the uniquely low-energy isomeric transition of $\Th$  at $\hbar\omega_{\rm iso}=8.4$~eV $(\lambda=148~{\rm nm})$ in $\Th$-doped fluoride crystals~\cite{tiedau_laser_2024,elwell_laser_2024} has enabled the first demonstrations of nuclear clocks~\cite{de2026thorium,huang2026nuclear}, leading to new frontiers in precision metrology and tests of fundamental physics~\cite{peik_nuclear_2003,beeks2021thorium,safronova2019search,fadeev2020sensitivity,beeks2025fine}. Besides, $\Th$-doped whispering-gallery-mode (WGM) resonators~\cite{kraemer2026toward} have been demonstrated, laying the groundwork for cavity QED with $\Th$. These advances establish solid-state $\Th$ as a timely platform at the interface of many-body quantum optics and frequency metrology. However, the nuclear coherence time in the host crystal is orders of magnitude shorter than the radiative lifetime~\cite{ooi2026frequency}, while the absence of a suitable auxiliary state prevents efficient repumping and state initialization. Thus, conventional Ramsey- and Rabi-type protocols are poorly suited to this regime~\cite{tiedau_laser_2024,elwell_laser_2024,morawetz_2026,kazakov_performance_2012}.

Existing nuclear clock implementations stabilize the interrogation laser frequency based on nuclear absorption spectroscopy~\cite{de2026thorium,huang2026nuclear}, whereas earlier proposals relied on fluorescence detection~\cite{kazakov_performance_2012}. The latter entails long cycle durations set by the isomer’s long radiative lifetime~\cite{kazakov_performance_2012}. Moreover, both approaches require a narrow-linewidth laser at $148$~nm, which remains a substantial technological challenge~\cite{lal2025continuous,xiao2026continuous, morawetz_2026,subramanian2026generation}. These limitations motivate collective interrogation schemes that harness the intrinsic metrological advantages of the nuclear transition while mitigating the constraints imposed by its solid-state implementation.

Here, we propose and analyze collective cavity QED-enhanced interrogation schemes for a solid-state nuclear frequency reference based on two- and one-photon pumping. 
Our key idea is to extract the clock error signal from the collective nuclear coherence through the resonator field, rather than from the slowly relaxing nuclear excitation. The next interrogation cycle can therefore begin once the coherence has decayed, without having to wait for radiative deexcitation.
We thus take advantage of what is typically a key limitation of solid-state emitters -- the large mismatch between the short coherence time and the long radiative lifetime -- to reset the clock on the coherence timescale.
In all three schemes, collective cooperativity governs the clock's fractional frequency instability. In the low-saturation regime, the instability is minimized when collective cooperativity equals unity, at which point the signal enhancement is balanced by induced cavity-mediated broadening. We further analyze the higher-saturation two-photon regime and identify a parameter interplay determining the optimum instability. While we focus on the specific case of $\Th$ in WGM resonators, our concept applies to a broader class of solid-state frequency standards. These results establish collective ensemble interrogation in nanophotonic resonators as a promising approach toward compact and stable solid-state nuclear frequency standards.
\begin{figure}
    \centering
\includegraphics[width=1\linewidth]{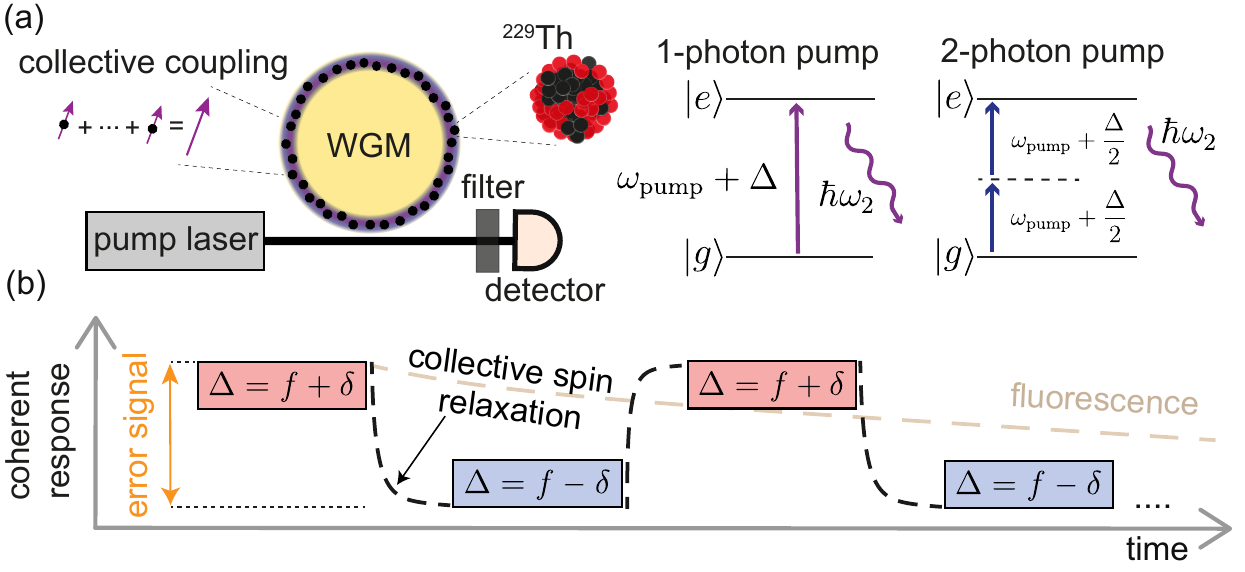}
    \caption{\textbf{Concept of a coherent nuclear frequency standard based on a whispering-gallery mode (WGM) resonator containing $\Th$.} (a) Schematic of the proposed setup and 1- or 2-photon pumping of the $\Th$ isomeric transition. (b) Schematic of the proposed interrogation sequences, leveraging the coherent transient signal to determine the offset of the pumping laser frequency from the nuclear transition, enabling fast state reinitialization and practically no dead time between cycles.}
    \label{fig:concept}
\end{figure}

\textit{$^{229}$Th in nanophotonic resonators}. In the following, we study a system of many ($N\gg1$) $^{229}$Th nuclei interacting with traveling-wave modes in a WGM resonator (see \cref{fig:concept}, and Ref.~\cite{kraemer2026toward} for experimental implementation).

We consider two nuclear excitation mechanisms.
One-photon pumping directly drives the resonant magnetic dipole~(M1) transition, as recently demonstrated experimentally \cite{tiedau_laser_2024}. Two-photon pumping exploits the recently predicted optonuclear quadrupolar (ONQ) effect \cite{xu2023two,xu2023solid}, which could enable interrogation at $296$~nm rather than $148$~nm through the second-order nuclear quadrupolar response to the electric-field gradient, enabling the use of commercially available laser systems, reducing crystal damage, and allowing operation in air rather than~vacuum. 

To enable homogeneous coupling of nuclei to the field in our two-photon-pumped case, we assume phase-matching between the two traveling-wave modes in the WGM resonator -- mode 1 at 296~nm wavelength ($\omega_1=\omega_{\textrm{iso}}/2$) and mode 2 at~148 nm~($\omega_2=\omega_{\textrm{iso}}$). This can be achieved, e.g., by resonator geometry tuning~\cite{NaturalWGM} or patterned implantation of the emitters \cite{pak2022long, jia2022integrated}. Thus, our system resembles collective cavity QED setups where atoms are engineered to couple homogeneously to the field, e.g., by placing them at the antinodes of a standing wave~\cite{Takamoto2005}. A detailed discussion on homogeneous coupling for the two-photon pumping case can be found in the Supplementary Information (SI), Section~S1. In the one-photon pumping (traveling-wave mode) case, such coupling emerges automatically (see SI, Section~S4). 

Under these assumptions, we can describe the interaction of the $\Th$ nuclei (modeled as two-level systems at the isomer transition frequency $\omega_{\textrm{iso}}$) with the photonic mode electromagnetic fields using the following Hamiltonian:
\begin{multline}
\hat{H}_S/\hbar =\Delta \hat{a}_2^{\dag}\hat{a}_2 
+\sum_{i}^{N} \Delta \frac{\hat{\sigma}^i_z} {2}+ \\+ 
\Big( G\hat{a}^{\dagger}_{2}\sum_{i}^{N}\hat{\sigma}_{-}^{i} + \Omega\sum_{i}^{N}\hat{\sigma}_{-}^{i}+{\mathcal{E}_2}\hat{a}^{\dagger}_{2} + \text{h.c.} \Big),   
\label{eq_mt:hamiltonian}
\end{multline}
where $\hat{a}_2$ (resp., $\hat{a}_2^{\dag}$) is the bosonic annihilation (creation) operator for photons in the mode 2 (at 148~nm), $\hat{\sigma}_{z}^{i} (\hat{\sigma}_{ee}^{i}=\frac{1+\hat{\sigma}_{z}^{i}}{2})$ and $
 \hat{\sigma}_{-}^{i} = (\hat{\sigma}_{+}^{i})^{\dagger}
$ are the standard spin inversion (excitation) and lowering operators for the $i$-th nucleus, and $\mathcal{E}_2$ is the direct one-photon pump at $148$~nm, used in the one-photon pumping scheme. 
The third term in~\cref{eq_mt:hamiltonian} is the interaction of mode 2 with nuclei via the M1 channel \cite{von_der_wense_theory_2020}, proportional to single-photon coupling strength $G=\frac{\mu}{\hbar} \sqrt{\frac{\mu_{0}\hbar\omega_{2}}{2V_2}}$, where $V_2$ is the magnetic field volume of the mode 2 and $\mu$ is the M1 transition matrix element.
The fourth term is the nuclear pumping through the ONQ effect with a two-photon Rabi frequency of $\Omega = \frac{1}{4}DE_1^2$, where $D$ contains the matrix element of the ONQ transition~\cite{kraemer2026toward, xu2023solid, xu2023two}, and $E_1^2=4P_1/(\varepsilon_0 n_1^2 V_1 \kappa_1)$, here $V_1,\kappa_1$, and $n_1$ being the field volume, cavity losses, and refractive index of mode 1, respectively; we assume $E_1$ is strong enough to be undepleted. 
Decay processes are described by standard jump operators in the Born-Markov master equation. These include cavity mode losses
with total decay rates comprised of intrinsic losses and external coupling: $\kappa_{1,2}=\kappa_{1,2}^{\rm int}+\kappa_{1,2}^{\rm ext}$ (hereafter critical coupling is assumed for both modes: $\kappa_{1,2}^{\rm int}=\kappa_{1,2}^{\rm ext}$); the spontaneous decay of nuclei at rate $\gamma$ and pure dephasing at $\gamma_{z}$ (see the details in the SI, Section~S1).

Furthermore, we focus on the case of CaF$_2$ as the host crystal, in which the nuclear spontaneous decay rate was measured: $\gamma = 1.59\times 10^{-3} \textrm{ s$^{-1}$}$~\cite{tiedau_laser_2024} (assuming only radiative decay). Hereafter, for simplicity we consider only the $|I = 5/2, m = \pm 1/2\rangle$ to  $|I = 3/2, m = \pm 1/2 \rangle$ sublevel transition (as the driving laser linewidth is typically much smaller than the quadrupole splitting)~\cite{beeks2021thorium, morawetz_2026}. Pure dephasing $\gamma_z$ is taken to be the nuclear spin relaxation rate due to interaction with surrounding fluorine magnetic spins in the crystal, unless stated otherwise: $\gamma_z =2\pi \times 300 \rm~Hz$ (other noise sources are assumed to be suppressed) \cite{kazakov_performance_2012}. Fluoride resonators with optical $Q$-factors exceeding $10^8$ have been demonstrated in the literature before~\cite{grudininfundamental2007, savchenkovselfinjection2019}; assuming weak degradation with the presence of $\Th$ dopant, we adopt an estimate of both $Q_{1,2}$ factors on the order of $10^6$, leading to cavity losses of $\kappa_{1,2} \sim 10^{10} \text{ } \rm { }s^{-1}$. WGM mode field volumes are taken to be $V_1(V_2)=2\times 10^{-13} \text{ } \rm m^{3}$ ($1\times  10^{-13}  \text{ } \rm m^{3}$), respectively~\cite{kraemer2026toward}, giving $G=20.4~\rm { }s^{-1}$ (thus, the single-emitter Purcell rate  $\sim G^2/\kappa_2 \ll \gamma$). A $\Th$ density of $\rho_{\rm Th}=6\times10^{24} \mathrm{\textrm{ } m^{-3}}$ \cite{morawetz_2026, tiedau_laser_2024} is assumed unless stated otherwise; the number of nuclei interacting with both modes is assumed to be $N = \rho_{\rm Th} V_2/3\equiv \rho V_2$ (because of density distribution over energy sublevels), resulting in $\kappa_2 \gg 4G^2N/\kappa_2$, which situates us in the so-called \enquote{bad-cavity} regime~\cite{walls2008quantum}.

\textit{Two-photon-pumped clock.} 
\begin{figure}[b]
    \centering
\includegraphics[width=1\linewidth]{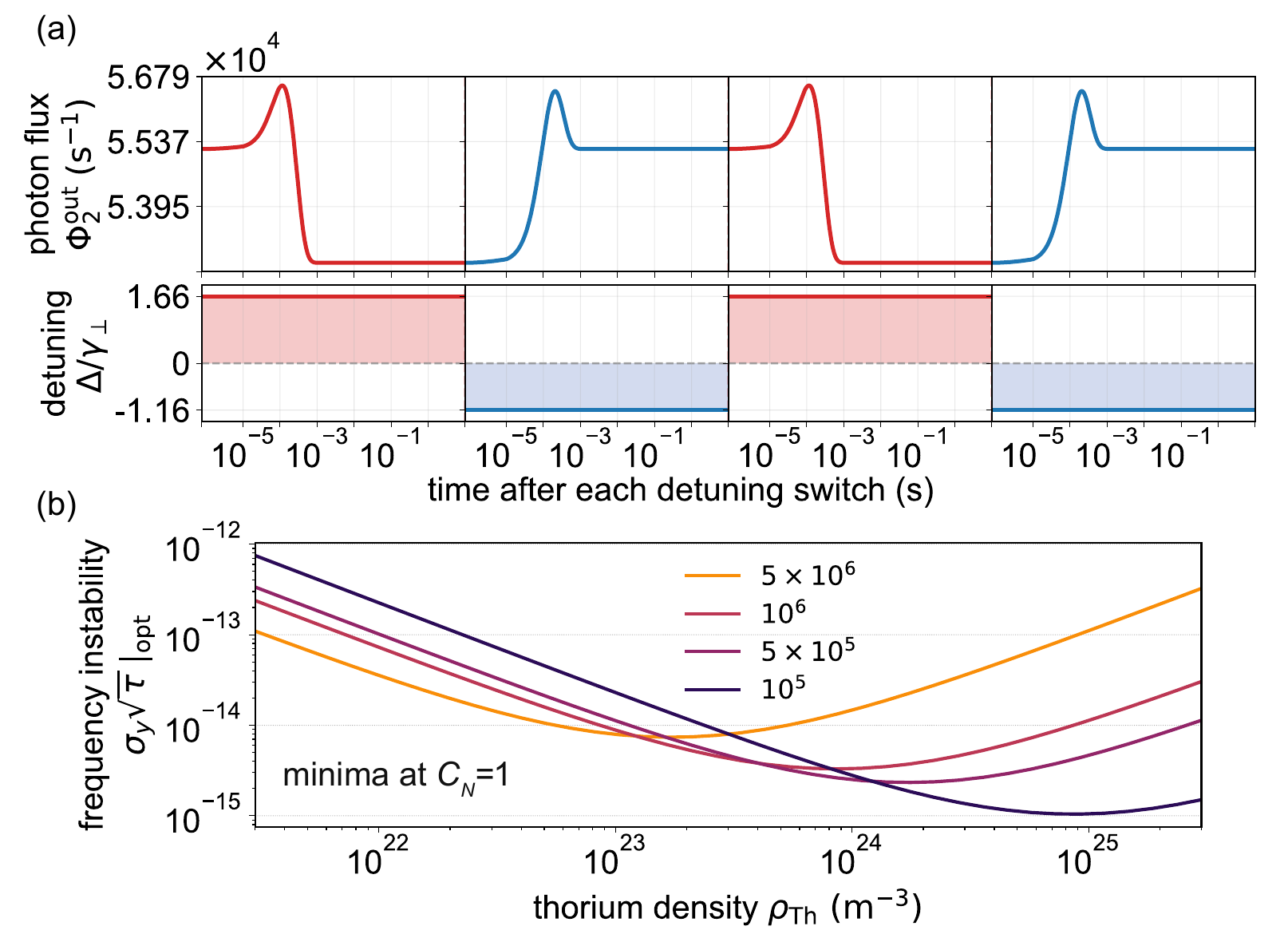}
    \caption{\textbf{Low-saturation steady state clock}. (a) Output photon flux $\Phi^\text{out}_2 = \kappa^\text{ext}_2 \langle a^{\dagger}_2 a_2 \rangle$ at $Q_1=5\times 10^6$, $Q_2=Q_1/8$ during two clock cycles with exemplary values of $f=\gamma_z/4,\delta =\sqrt{2}\gamma_z$ (for illustration). (b) Optimized fractional frequency instability of the clock  $\left. \sigma_y\sqrt{\tau} \right|_{\mathrm{opt}}$ as a function of $\rho_{\rm Th}$ for several $Q_2$-factors, at fixed $Q_1 = 5\times 10^6$. For both plots $P_1 = 0.4$~mW.}
    \label{fig:four-detuning}
\end{figure}
In this case, \cref{eq_mt:hamiltonian} is simplified by $\mathcal{E}_2=0$, and $\Delta = \omega_{2}-2\omega_{\rm pump,\text{ }  296 \text{ } nm}$. We use the second- and first-order (i.e., mean-field) cumulant expansion ~\cite{Plankensteiner2022, kubo1962generalized} to find the average values of system observables (see SI, Sections~S2 and S5).
\begin{figure*}[!htb]
    \centering
    \includegraphics[width=1.0\linewidth]{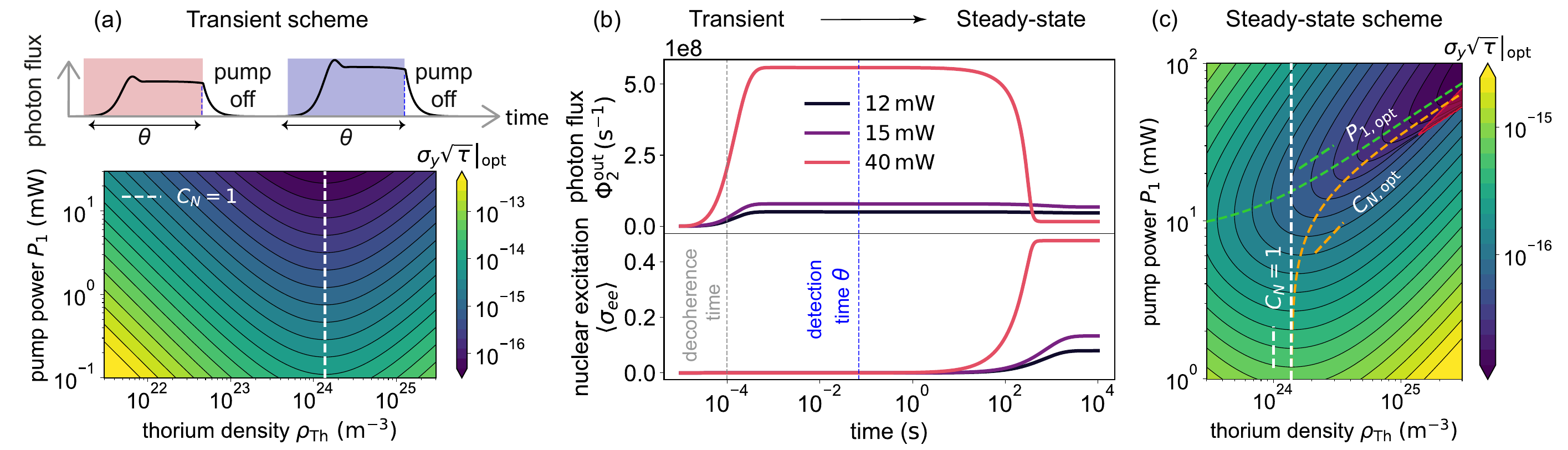}
    \caption{\textbf{From transient to steady-state frequency standards}. (a) Transient frequency standard interrogation scheme and corresponding optimized frequency instability as a function of $\rho_{\rm Th}$ and $P_1$. (b) Output photon flux dynamics $\Phi_2^{\rm out}$ and nuclear excitation $\sigma_{ee}$ for three pump powers, featuring the transient pulse-like behavior and the relaxation towards the steady state. (c) General frequency instability as a function of pump power $P_1$, thorium density $\rho_{\rm Th}$, at each point separately numerically optimized with respect to $\delta$. All plots are at fixed $Q_1=5\times 10^6, Q_2=Q_1/8$. The red region on the colormap shows the bistability range predicted by mean-field analysis.}
    \label{fig:2ph_instability}
\end{figure*}
In the bad-cavity regime, $\hat{a}_2$ adiabatically follows the nuclear coherence: $\hat{a}_2(t) \simeq -2iG\kappa_{2}^{-1}\sum^N_{i}\hat{\sigma}_{-}^{i}(t)$, and the photon number is (dropping hat signs):
\begin{align}
    n_2(t) & \approx 4G^{2}\kappa_{2}^{-2}\left[N\langle \sigma_{ee}\rangle (t)+N(N-1)\langle\sigma_{+}^{1}\sigma_{-}^{2}\rangle(t)\right]\notag \\
    &\equiv n_{2}^\text{ind}(t)+n_{2}^\text{coll}(t), \label{eq:coll_ind}    
\end{align}
decomposed into a sum of independent ($\propto N\langle \sigma_{ee}\rangle$) and collective ($\propto N(N-1)\langle\sigma_{+}^{1}\sigma_{-}^{2}\rangle$) emission contributions, where $\langle\sigma_{+}^{1}\sigma_{-}^{2}\rangle$ are pairwise internuclear correlations.
We find that the system dynamics are captured correctly by the mean-field equations in the relevant parameter ranges, which read (assuming all nuclei are identical):
\begin{equation}
\left\{
\begin{aligned}
\langle \dot{\sigma}_{-}\rangle
&= -\left[i\Delta+\gamma_{\perp}
-\frac{2G^{2}N}{\kappa_2}
(2\langle\sigma_{ee}\rangle-1)
\right]
\langle\sigma_-\rangle \\[-2pt]
&\qquad +2i\Omega\langle\sigma_{ee}\rangle-i\Omega, \\[4pt]
\langle \dot{\sigma}_{ee}\rangle &=
-\frac{4G^{2}N}{\kappa_2}
|\langle\sigma_-\rangle|^{2}
+i\Omega \left(\langle\sigma_-\rangle-\langle\sigma_-\rangle^{*}
\right) -\gamma\langle\sigma_{ee}\rangle,
\end{aligned}
\right.
\label{eq_mt:meanfield_system}
\end{equation}
where $\gamma_{\perp}=\gamma/2+\gamma_{z}$ ($\gamma_{\perp} \approx \gamma_z$), such that the coherent field relaxation rate is dominated by the sum of pure nuclear dephasing $\gamma_z$ and collective cavity-induced decoherence rate $2G^2N/\kappa_2$ ($\gamma_{z}, 2G^{2}N/\kappa_{2} \gg \gamma $) near ground state. Furthermore, we use the collective cooperativity parameter $C_N$ that describes the strength of collective light-matter interactions: $C_N
= 2 G^2 N/(\kappa_2\gamma_\perp)$~\cite{plankensteiner2019enhanced}. 

The performance of a frequency standard is described by its fractional frequency instability $\sigma_y$, defined as the fractional frequency error $\sigma_f/\omega_{\rm iso}$ in determining the angular frequency offset $f$ between the laser and transition frequencies (which is then to be corrected via a feedback loop) in one cycle, averaged over $k$ cycles (assuming uncorrelated noise): $\sigma_y  =\frac{\sigma_{f}}{\omega_{\rm iso}} \frac{1}{\sqrt{k}}= \frac{\sigma_{f}}{\omega_{\rm iso}} \frac{\sqrt{T_c}}{\sqrt{\tau}}$. Here, $T_c = 2\theta$ is the duration of one interrogation cycle, and $\tau$ is the total averaging duration. 

The procedure of (periodically) extracting the offset $f$ and accordingly adjusting the laser frequency $\omega_{\rm pump}$ constitutes the interrogation scheme. In what follows, each cycle employs a two-point sampling scheme: in the first half-cycle, the laser frequency is red-detuned at $\Delta_{+} = f+\delta$ and the photons at the cavity output are detected; then the laser is switched to blue detuning at $\Delta_-=f-\delta$ and similar detection occurs; here $\delta$ is an adjustable working point. The difference between these two photon counts provides the error signal for estimating $f$, after which the cycle is repeated continuously (see \cref{fig:concept}(b)).
The central idea underlying all our proposed approaches is to read out the \textit{collective} coherent optical signal from the cavity, which switches between the quasi-equilibria at $\Delta_{\pm}$ \textit{on the decoherence timescale rather than the (much slower) population lifetime.} This enables rapid, continuous state reinitialization and readout almost without dead time~\cite{kazakov_performance_2012}, thus suppressing the so-called Dick effect~\cite{quessada2003dick}.

In the steady-state low saturation case ($\sigma_{ee} \approx 0$), the collective radiation dominates the photon flux: $\Phi^\text{out}_2(\Delta_{\pm}=f\pm \delta)=\frac{\kappa_2}{2}n_{2}(\Delta_{\pm}) $. Two full cycles of switching $\Delta$ are shown in \cref{fig:four-detuning}(a), where we see an initial transient spike followed by an established quasi-equilibrium (the former yielding negligible photon count -- note the logarithmic time axis). Thus, the difference $\Delta N_{\rm det} \simeq [\Phi^\text{out}_2(f +\delta) - \Phi^\text{out}_2(f -\delta)] \times \theta $ contains information about the offset $f$, while $\delta$ is a parameter that should be optimized to yield minimal instability. Analytically deriving and substituting the expression for the optimal $\delta_{\rm opt}=\frac{\gamma_{\perp}(1+C_N)}{\sqrt{2}}$, we find the corresponding optimal $\left.
\sigma_y\sqrt{\tau}
\right|_{\mathrm{opt}}$ in the shot-noise limit~(see SI, Section~S3~A):
\begin{align}
\sigma_y\sqrt{\tau}|_{\mathrm{opt}}
=\frac{3\sqrt{6}}{4} \frac{G\gamma_\perp}{\omega_{\mathrm{iso}}\Omega\sqrt{\kappa_2}}\frac{(1+C_N)^2}{C_N}.
\label{eq_mt:optimal_inst}
\end{align}

Remarkably, the optimal fractional frequency instability features a minimum corresponding to $C_N = 1$ (equivalently, $\rho =\kappa_2\gamma_{\perp}/(2G^2V_2)$), as shown in \cref{fig:four-detuning}(b).

This result has the following intuitive physical explanation: at low $\Th$ density $\rho$, increasing the number of nuclei $N$ enhances the collective signal-to-noise ratio ($\propto N$), while collective cavity-mediated broadening remains negligible because pure dephasing dominates, $\gamma_{\perp} \gg 2G^2N/\kappa_2$. The clock instability therefore decreases with increasing $\rho$ until $C_N=1$. Beyond this point, the cavity-mediated term dominates $\gamma_{\perp} \ll 2G^2N/\kappa_2$, such that further increasing $\rho$ enhances decoherence and worsens the instability. Two practical corollaries follow. First, a higher cavity quality factor $Q_2$ is not always beneficial and may even degrade clock performance. Second, additional cavity losses -- for example, from implantation-induced resonator damage -- do not necessarily compromise the achievable stability, which is encouraging for future experiments. Notably, the instability (at $\tau =1$~s) reaches $\sim10^{-17}$ levels, showcasing performance similar to the fluorescence interrogation scheme analyzed in~\cite{kazakov_performance_2012} at the same value of dephasing $\gamma_z$.

\cref{eq_mt:optimal_inst} suggests a simple route to decrease $\sigma_y\sqrt{\tau}$ -- that is, to increase pump strength $\Omega$.
At sufficiently high pump strengths, however, significant nuclear excitation invalidates the preceding analysis.
To reveal the nature of system dynamics, we plot the evolution of output photon flux $\Phi^\text{out}_2$ and nuclear excitation $\sigma_{ee}$ to the steady state for several exemplary pump powers in~\cref{fig:2ph_instability}(b), described by \cref{eq_mt:meanfield_system}. The first timescale of $\Phi^\text{out}_2$ dynamics is set by the total decoherence rate $\gamma_{\perp}(1 + C_N)$ and does not depend on pump power, leading to an intermediate plateau photon number $n^{\rm plat}_2 \propto \Omega^2$ (essentially describing a driven-dissipative oscillator). The second timescale is governed by the nuclear excitation and spontaneous decay rates (see SI, Section~S2 A), and becomes shorter as the pump power is increased.

Two further collective readout protocols emerge naturally. The first one is to measure the transient optical signal before significant nuclear excitation, during time $0<t<\theta$, resulting in approximately
$N_{\text{det}}(\Delta,\theta ) = \kappa_{2}^{\text{ext}} n^{\rm plat}_{2}(\Delta)\times \theta $. It is clear from \cref{fig:2ph_instability}(b) that the detection time $\theta$ can be adjusted within a reasonable range between approximately $0.01-0.1$~s $\ll 1 /\gamma$, enough to collect a significant number of photons. Then, the pumping laser is turned off, and -- crucially -- one again has to wait only for the coherence to decay (and thus the cavity field, adiabatically slaved to it) to reinitialize the state of the system, since the nuclei are essentially not excited yet (see \cref{fig:2ph_instability}(a)). Repeating the same steps with the other laser detuning, we can thus determine $f$. The next cycle may commence essentially instantly (compared to $1 /\gamma$); the mathematical description of this \enquote{transient} scheme is identical to the steady-state low-saturation case (since we operate while $\sigma_z\approx -1 $), provided that $\theta$ is decreased accordingly to the increase in pump power. However, inevitably, a small portion of the nuclei will be excited at the end of each such half-cycle. Over many cycles, this parasitic excitation will build up and eventually necessitate a \enquote{reset} period lasting on the order of a radiative lifetime. This limits the viability of the transient scheme for long-term clock operation; however, it can be used as an auxiliary intermittent clock that is run for short periods of time along with a flywheel oscillator to generate time scales~\cite{hachisu2018months, yao2019optical}.

The second protocol is to drive the system into the steady state, and then start operating the above red-blue-detuned pumping scheme. Even at high saturation, the photon count will follow the nuclear coherence, whereas the nuclear excitation level will not change significantly across cycles of laser frequency changes, provided that the cycles (and thus $\theta$) are short enough~(which is possible in the parameter regimes considered since the pumping rates remain well below the decoherence rates). After each detuning switch, the coherence therefore relaxes to a quasi-steady value, while the population does not have sufficient time to reach the new true steady state. Thus, by “steady-state scheme” we mean that the protocol is initiated after the system has first reached steady state, whereas during the subsequent detuning change cycles only a quasi-steady state is established, allowing us to extract the information about laser detuning. \cref{fig:2ph_instability}(b) shows that the steady state photon count depends on the pump power in a non-monotonous manner. Hence, as opposed to the transient scheme case, a more mathematically involved co-optimization of $P_1$ and $\delta$ is needed to minimize the clock instability (see SI, Section~S3 for extended analytical derivations). 

We therefore plot $\left. \sigma_y\sqrt{\tau}\right|_{\mathrm{opt}} (P_1, \rho)$ in \cref{fig:2ph_instability}(c), obtained by numerically optimizing the working point $\delta$ at each $(P_1, \rho)$ (implying that $\delta= \delta_{\mathrm{opt}} (P_1,\rho)$). For a fixed collective cooperativity $C_N$, we can analytically determine the optimal pump power $P_{1, \rm opt}$ and $\delta_{\rm opt}$ separately (see SI, Section~S3 C1). 

The analytical prediction for $P_{1,\mathrm{opt}}$, overlaid as the dashed green line, closely follows the numerically calculated minima indicated by the colormap contours in \cref{fig:2ph_instability}(c).
Optimizing instead over $C_{N}$ at fixed $P_1$ and an optimized detuning $\delta_{\rm opt}$ yields $C_{N, \rm opt} = 1+\frac{4\Omega^2}{6\gamma \gamma_{\perp}}$, shown by the orange dashed line. This expression recovers the low-saturation optimum $C_N \to 1$ as $\Omega \to 0$ (SI, Section~S3 C2). At these optimized operating conditions, the predicted fractional instability reaches the $10^{-17}$ level at $\tau=1$~s.

Importantly, the mean-field analysis reveals a bistability region at large pump powers (shown in red in \cref{fig:2ph_instability}(c)). In this regime, the steady-state equation for $\sigma_z$ admits three solutions: two stable and one unstable. Near the bistable region, $\sigma_y$ reaches a minimum and exhibits a discontinuity. This discontinuity arises because, at each point in the colormap, we select the stable solution for $\sigma_z$ that yields the lowest instability; the optimum therefore switches between the two stable branches. Consequently, the optimal solution may lie on the branch with larger $\sigma_z$, corresponding to a more strongly excited nuclear ensemble.
If bistability is observed experimentally, accessing the desired high-$\sigma_z$ branch will require deliberate state preparation, for example through a strong pump pulse applied before clock operation. Otherwise, an initially unexcited ensemble is expected to settle onto the lower-$\sigma_z$ branch. Whether the predicted bistability represents genuine nonlinear dynamics---as in refractive optical bistability~\cite{boyd2008nonlinear,soljavcic2004enhancement} or few-atom cavity-QED systems~\cite{dombi2013optical}---or instead arises from the mean-field approximation~\cite{mendoza2016beyond} remains an open question for future experiments. Unlike the transient protocol, this steady-state scheme can, in principle, operate continuously and is therefore viable for a practical clock.

\begin{figure}[b]
    \centering
\includegraphics[width=1\linewidth]{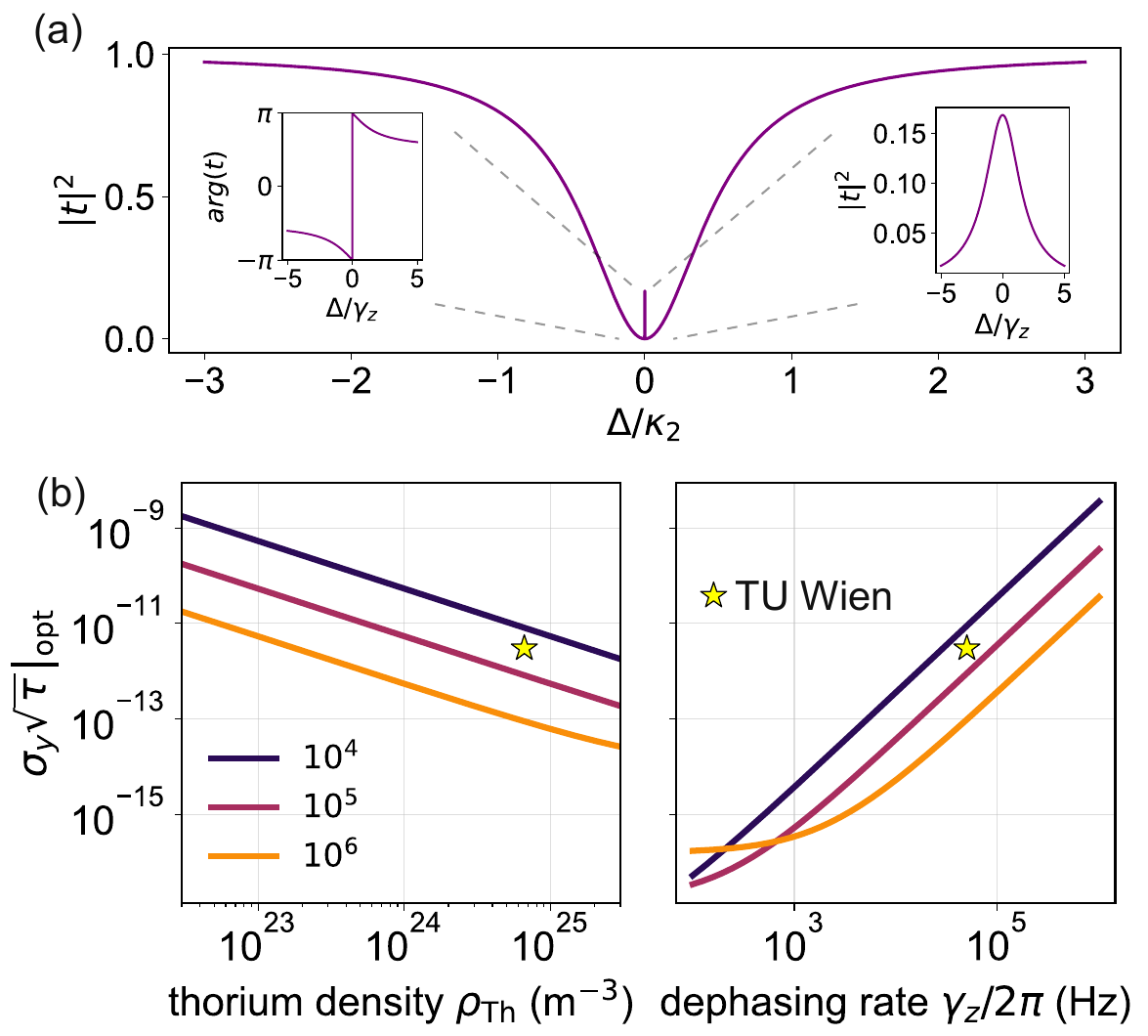}
    \caption{\textbf{Nuclear clock based on collective cavity antiresonance.} (a) Antiresonance in transmission as a function of $\Delta$: the nuclei introduce a phase shift around the resonance, thus breaking the destructive interference condition between the outcoupled light from the resonator and the transmitted light from the input waveguide. (b) Comparison of frequency instability of the antiresonant clock with the TU Wien nuclear clock experiment~\cite{de2026thorium} at $P_2=100$~pW pump power, $\gamma_z = 2\pi \times 50$~kHz, and $\rho_{\rm Th}=6 \times 10^{24} \text{ } \rm {m^{-3}}$. The different lines correspond to $Q_2 =10^4,10^5,10^6$.}
    \label{fig:1ph_clock}
\end{figure}
\textit{Antiresonant clock}. While the two-photon pumping process based on ONQ is a promising theoretical proposal, to this date only one-photon driving was demonstrated experimentally (both in fluorescence and absorption spectroscopy)~\cite{morawetz_2026, tiedau_laser_2024}. This motivates the development of a one-photon interrogation scheme that benefits from collective cavity-QED effects. In this case, $\mathcal{E}_2 \neq 0$, $\Omega = 0$, and $\Delta = \omega_2-\omega_{\rm pump, \text{ } 148 \text{ }  nm}$ in \cref{eq_mt:hamiltonian}.
We consider the low-saturation (linear) regime in this case since the available 148 nm laser sources currently yield  small powers. Using input-output relations, the steady-state cavity transmission is (see SI, Section~S4):
\begin{equation}
t= -1+\frac{\kappa_{2}}{2}\frac{\left(\gamma_{\perp}+i\Delta\right)}{(\frac{\kappa_{2}}{2}+i\Delta)\left(\gamma_{\perp}+i\Delta\right)+G^{2}N}.
\end{equation}
This transmission function features a narrow peak on top of the cavity linewidth, a phenomenon usually referred to as \enquote{cavity antiresonance}~\cite{plankensteiner2017cavity, sames2014antiresonance, alsing1992suppression, zippilli2004suppression} (see \cref{fig:1ph_clock}(a)). 
Notably, in this case the homogeneous collective coupling emerges naturally, bypassing the need for phase-matching engineering between two cavity modes. 
Deriving $\sigma_y \sqrt{\tau}$ in the same two subsequently red-blue detuned laser interrogation scheme that similarly features state reinitialization on the decoherence timescale, we find $\left. \sigma_y\sqrt{\tau}\right\rvert_{\mathrm{opt}}
=
\frac{3\sqrt{3}}{4}
\frac{ \gamma_{\perp}}{\omega_{\rm iso}}
\frac{(1+C_N)^{2}}{C_N}
\sqrt{
\frac{\hbar\omega_{\rm iso}}{P_2}
}$, such that again instability is minimized when $C_N=1$.

To showcase the potential of the proposed WGM resonator approach, we compare it to a recent nuclear clock experiment with a feedback loop at TU Wien~\cite{de2026thorium}, where a Th:CaF$_2$ crystal with a thorium density
of $\rho_{\rm Th} = 6.6 \times 10^{24} \text{ } \rm m^{-3}$ was used, and $\sigma_y(\tau=1\text{ } \mathrm{s})=3 \times 10^{-12}$ has been demonstrated~\cite{de2026thorium}.
We now adopt a realistic linewidth of the nuclear transition (FWHM of $100$~kHz, meaning $\gamma_z = 2\pi \times 50$~kHz), used in that work, and a similar pump power of $P_2=100$~pW at $148$~nm wavelength. We see in \cref{fig:1ph_clock}(b) that with reasonable $Q_2$-factors, we may expect frequency instability comparable to that of the demonstrated absorption spectroscopy-based clock. For another reference, the nuclear clock at Tsinghua University~\cite{huang2026nuclear} also shows similar values ($\sigma_y(\tau=1\text{ } \mathrm{s})=2 \times 10^{-12}$); however, in that work a much higher pump power was used and a narrower linewidth reported (FWHM of $\sim 30$~kHz). Hence, overall these numbers are intended to indicate the potentially achievable performance rather than to provide a rigorous quantitative comparison.
Most notably, the total number of nuclei interacting with the mode in our design ($N\sim 10^{11}$, provided that $\Th$ is implanted mostly into the region of electromagnetic mode field~\cite{kraemer2026toward}) is many orders of magnitude smaller than the amount of $\Th$ interacting with the laser beam in the mentioned TU Wien nuclear clock based on a Th:CaF$_2$ crystal of approximately the same doping density~($N\sim10^{16}$). Thus, the WGM resonator and cavity-QED enhancement employed in the proposed scheme could drastically reduce the required quantity of $\Th$, an important advantage given its scarcity and difficulty of production.

To conclude, we have proposed and analyzed three collective cavity-QED-enhanced optical clock interrogation schemes, building upon the recent breakthroughs on the solid-state $\Th$ nuclear clock and $\Th$ WGM resonator fabrication as a promising platform~\cite{de2026thorium, huang2026nuclear, kraemer2026toward}: a two-photon nuclear clock, a two-photon transient frequency standard, and a one-photon antiresonant clock. All schemes leverage the collective coupling of nuclei to the photonic modes of the resonator and use the drastic mismatch between the nuclear coherence and population lifetimes to operate the clock with much faster state initialization.
We demonstrate that these schemes can yield great performance, while requiring orders of magnitude fewer $\Th$~nuclei. 
We further derive closed-form design rules linking the clock instability to collective cooperativity, pump power, and detuning, including the universal low-saturation optimum $C_N=1$ and its extension to the saturated regime.
Our work paves the way for solid-state cavity QED-enhanced optical metrology, and motivates further experimental research in this direction. 

\section*{Competing interests}
TS and CRC are seeking patent protection for ideas in this work (application number EP25227524.3).

\section*{Data and code availability statement}
The data and code that support the plots within this paper and other findings of this study are available from the corresponding author upon reasonable request. 

\section*{Acknowledgments}
The authors acknowledge discussions with Haowei Xu, Ju Li, Chen Mechel, Aliaksei Horlach, Aviv Karnieli, Ido Kaminer, Kjeld Beeks, Swadheen Dubey, Jan de Haan, Kasper van Gasse, and Jamison Sloan. 
\section*{Funding}
This research was funded in part by the Austrian Science Fund (FWF) 10.55776/KIN2703425. For open access purposes, the author has applied a CC BY public copyright license to any author-accepted manuscript version arising from this submission. CRC acknowledges startup funding from the Institute of Science and Technology Austria (ISTA). 

\bibliography{bibliography}

\clearpage
\includepdf[pages={{},-}]{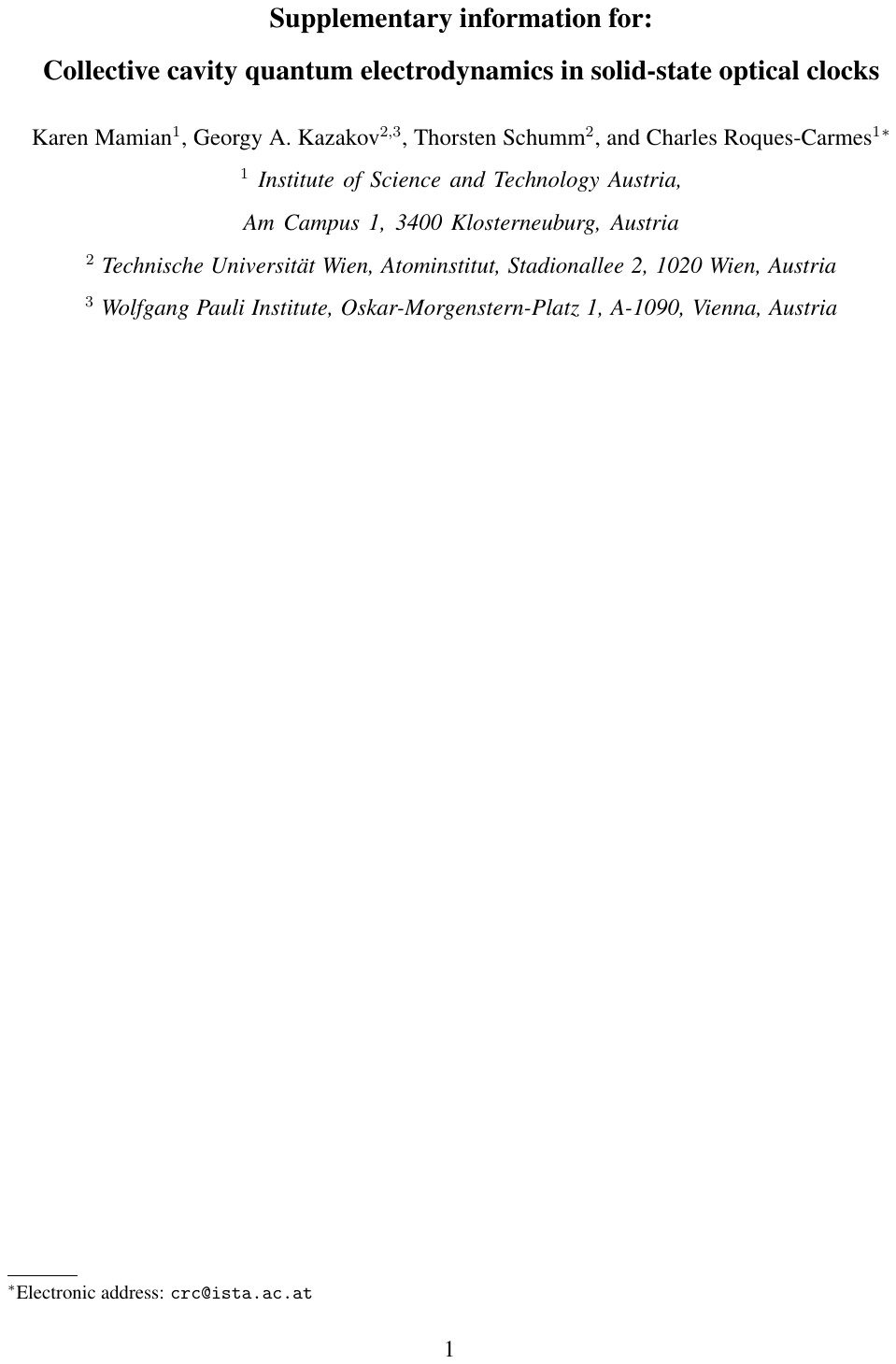}
\AddToHookNext{shipout/before}{\DiscardShipoutBox}
\end{document}